\documentstyle[12pt,a4,cite,epsfig]{article}

\newcommand{\ba}{\begin{eqnarray}}
\newcommand{\ea}{\end{eqnarray}}
\newcommand{\be}{\begin{equation}}
\newcommand{\ee}{\end{equation}}
\newcommand{\dis}{\displaystyle}

\begin{document}

\begin{titlepage}
\begin{flushright}
CAFPE-97/08\\
UG-FT-227/08
\end{flushright}
\vspace{2cm}
\begin{center}

{\large\bf  Hadronic Light-by-Light Contribution to Muon
$g-2$: Status and Prospects
\footnote{Invited talk at ``PHIPSI08, 
International Workshop on $e^+e^-$ collisions from Phi to Psi'', 
April 7-10  2008, Frascati, Italy.}}\\
\vfill
{\bf  Joaquim Prades}\\[0.5cm]
CAFPE and Departamento de
 F\'{\i}sica Te\'orica y del Cosmos, Universidad de Granada, 
Campus de Fuente Nueva, E-18002 Granada, Spain.\\[0.5cm]

\end{center}
\vfill

\begin{abstract}
\noindent
I review the recent calculations and present status of the
hadronic light-by-light contribution to muon $g-2$.
\end{abstract}
\vfill
June 2008
\end{titlepage}
\setcounter{page}{1}
\setcounter{footnote}{0}

\section{Introduction}
\begin{figure}[htb]
\label{fig1}
\begin{center}
\epsfig{file=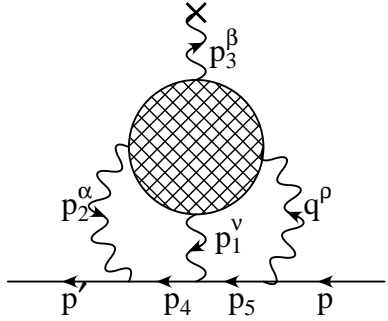,width=5cm}
\end{center}
\caption{One momenta configuration of the hadronic light-by-light
contribution to muon $g-2$.}
\end{figure}

Here, I discuss the contribution to the muon $g-2$ of a hadronic 
bubble connected to the external static magnetic 
source through one photon leg and to the muon line with  
another three photon legs. This corresponds to the so-called hadronic
light-by-light contribution to the muon anomaly $a_\mu= (g-2)/2$. 
Recent reviews are in \cite{BP07,PRV08}. One of the six
possible photon momenta configurations is shown in Fig. \ref{fig1}
 and its contribution to the  vertex
$-|e| \, \bar{u}(p^\prime) \, \Gamma^\beta (p-p^\prime) \, 
u(p) \, A_\beta$ is 
\ba
\label{Mlbl}
\dis{\Gamma^\beta} (p_3)
&=&  - e^6  
\int {{\rm d}^4 p_1 \over (2\pi )^4}
\int {{\rm d}^4p_2\over (2\pi )^4}  
{\Pi^{\rho\nu\alpha\beta} (p_1,p_2,p_3) 
\over q^2\, p_1^2 \, p_2^2}  
\gamma_\alpha (\not{\! p}_4-m )^{-1} 
\gamma_\nu (\not{\! p}_5 -m )^{-1} \gamma_\rho \, 
 \nonumber \\ 
\ea
where $p_3 \to 0 $ is the momentum of the
photon that couples to the external magnetic source,
$q=p_1+p_2+p_3$ and $m$ is the muon mass. 
 The dominant contribution to the hadronic four-point  
function 
\ba
\label{four}
\Pi^{\rho\nu\alpha\beta}(p_1,p_2,p_3)&=& \nonumber \\
i^3 \int {\rm d}^4 x \int {\rm d}^4 y
\int {\rm d}^4 z \, {\rm e}^{i (p_1 \cdot x + p_2 \cdot y + p_3 \cdot z)}
 \langle 0 | T \left[  V^\rho(0) V^\nu(x) V^\alpha(y) V^\beta(z)
\right] |0\rangle && 
\ea
comes from the three light quark 
$(q = u,d,s)$ components in the electromagnetic current
$V^\mu(x)=\left[ \overline q \widehat Q \gamma^\mu q \right](x)$
with $\widehat Q$ the quark electrical  charge matrix. 

Using  gauge-invariance, one can write
\be
\Pi^{\rho\nu\alpha\beta}(p_1,p_2,p_3)
= - p_{3 \lambda} \, \frac{\partial \Pi^{\rho\nu\alpha\lambda}
(p_1,p_2,p_3)}  {\partial p_{3 \beta}} \Big|_{p_3=0}\, 
\ee
 and therefore one just needs derivatives of the 
four-point function at $p_3=0$.  
 The contribution to $a_\mu$  is
\be
a_\mu^{\rm lbl}
=\frac{1}{48 m} {\rm tr}
\{ (\not{\! p}+m) [ \gamma^\beta , \gamma^\lambda ] (\not{\! p}+m) 
\frac{\partial \Gamma_\lambda(0)}
{\partial p_{3 \beta}} \} \, .  \, 
\ee
The four-point function $\Pi^{\rho\nu\alpha\beta}(p_1,p_2,p_3)$
is an extremely difficult object involving  many scales and  
no full first principle calculation of it  has been reported yet.
Notice that we need momenta $p_1$ and $p_2$
 varying from 0 to $\infty$.
Unfortunately, there is neither  a direct connection of $a_\mu^{\rm lbl}$
 to a measurable quantity. Two lattice groups have started 
exploratory calculations \cite{HB06,RAK08}  but the 
final uncertainty that these calculations can reach is not clear yet.

Attending to a combined  large number of colors $N_c$  of QCD
 and chiral perturbation
theory  (CHPT) counting one can
distinguish four types of contributions \cite{EdR94}.
Notice that the CHPT counting is only for organization of the
contributions and refers to the lowest order term contributing
in each case. The four different types of contributions are:
\begin{itemize}
\item Goldstone boson exchanges contribution are ${\cal O}(N_c)$
and start  at ${\cal O}(p^6)$ in CHPT.
\item One-meson irreducible vertex contribution and 
non-Goldstone boson exchanges contribute  also at
${\cal O}(N_c)$  
but start contributing at ${\cal O} (p^8)$ in CHPT.
\item One-loop of Goldstone bosons contribution
 are ${\cal O}(1/N_c)$ and start at ${\cal O}(p^4)$ in CHPT.
\item One-loop of non-Goldstone boson contributions 
which are ${\cal O}(1/N_c)$  but start contributing  
at ${\cal O}(p^8)$ in CHPT.
\end{itemize}
Based on the counting above there are two full calculations
\cite{HKS98,HK02}  and \cite{BPP96,BPP02}. 
There is also a detailed study of the $\pi^0$
exchange contribution \cite{KN02} putting emphasis in obtaining
analytical expressions for this part.

Using  operator product expansion (OPE) in QCD, 
the authors of \cite{MV04} pointed out a new short-distance
 constraint of the reduced full four-point Green function
\ba
\langle 0 | T \left[ 
V^\nu (p_1) V^\alpha (p_2) V^\rho (-(p_1+p_2+p_3)) \right]
| \gamma(p_3) \rangle  
\ea
when $p_3 \to 0$  and in the special momenta configuration 
 $-p_1^2 \simeq -p_2^2 >> -(p_1+p_2)^2$ Euclidean and large.
See also \cite{KPM04}. This short distance constraint was not explicitly
imposed in previous calculations. 

\section{Leading in the $1/N_c$ Expansion Contribution} 
\label{LOC}

Using effective field theory techniques, the authors of
 \cite{KNP02} shown that leading contribution to $a_\mu^{\rm lbl}$
contains a term  enhanced by a $\log^2 (\mu/m)$ factor  
 where $\mu$ is an ultraviolet scale and the muon mass $m$ provides
the infrared scale.  This leading logarithm is generated 
by the Goldstone boson exchange
contributions and is fixed by the Wess--Zumino--Witten (WZW) 
vertex $\pi^0 \gamma \gamma$.  
In the chiral limit where quark
 masses are neglected and at large $N_c$, the coefficient of this 
double logarithm    is model  independent and has been calculated   
and shown to be  positive in \cite{KNP02}. 
All the calculations we discuss here 
agree with these leading behaviour  and its coefficient including the 
sign.  A global sign mistake in the $\pi^0$ exchange in \cite{HKS98,BPP96}
was found by \cite{KN02,KNP02} and confirmed by \cite{HK02,BPP02} 
and by others \cite{BCM02,RW02}. The subleading 
$\mu$-dependent terms \cite{KNP02}, namely, 
$\log(\mu/m)$ and a non-logarithmic term $\kappa(\mu)$, 
 are model dependent and calculations of them are implicit in the
results presented in  \cite{HKS98,HK02,BPP96,BPP02,MV04}. 
 In particular, $\kappa(\mu)$ contains the large $N_c$ contributions from 
the one-meson irreducible vertex and non-Goldstone boson exchanges.
In the next section we review the recent model calculations
of the full leading in the $1/N_c$ expansion contributions.

\subsection{Model Calculations}

 The $\pi^0$ exchange contribution was calculated
in  \cite{HKS98,HK02,BPP96,BPP02,KN02,DB08}  by 
constructing  the relevant four-point function 
in terms of the off-shell  $\pi^0 \gamma^*(p_1) \gamma^*(p_2)$
form factor ${\cal F}(p_1^2, p_2^2)$ and the off-shell
$\pi^0 \gamma^*(q) \gamma(p_3=0)$ form factor
${\cal F}(q^2, 0)$   modulating each a WZW $\pi^0 \gamma \gamma$ vertex. 
In all cases several short-distance QCD  constraints 
were imposed on these form-factors.
 In particular, they all have the correct QCD short-distance behaviour
\be
{\cal F}(Q^2, Q^2) \to \frac{A}{Q^2}
\hspace*{0.5cm} {\rm and} \hspace*{0.5cm} 
{\cal F}(Q^2, 0) \to \frac{B}{Q^2}
\ee
when $Q^2$ is Euclidean
 and are  in agreement with $\pi^0\gamma^*\gamma$ data.
\begin{table}
\begin{center}
\caption{Results for the $\pi^0$,
$\eta$ and $\eta'$ exchange contributions.
\label{tab1}}
{\begin{tabular}{c|cc}
 Reference &\multicolumn{2}{c}{ $10^{10} \times a_\mu$}\\
 & $\pi^0$ only &  $\pi^0$, $\eta$ and $\eta'$\\
\hline
\cite{HKS98,HK02}  & 5.7 & 8.3 $\pm$ 0.6 \\
\cite{BPP96,BPP02} & 5.6  & 8.5 $\pm$ 1.3 \\
\cite{KN02} with $h_2=0$ & 5.8 & 8.3 $\pm$ 1.2\\
\cite{KN02} with $h_2=-10$~GeV$^2$ & 6.3 & \\
\cite{DB08} & 6.3 $\sim$ 6.7 & \\
\cite{MV04} &  7.65 &11.4$\pm$1.0  
\end{tabular}}
\end{center}
\vspace*{-0.5cm}
\end{table}
 They  differ slightly in shape  due to the different model
assumptions (VMD, ENJL, Large $N_c$, N$\chi$QM) 
but they produce small numerical differences always compatible 
within quoted uncertainty $\sim 1 \times 10^{-10}$ 
--see Table \ref{tab1}.

Within the models used in \cite{HKS98,HK02,BPP96,BPP02,KN02,DB08}, 
to get the full contribution at leading  in $1/N_c$ 
one needs to add the one-meson irreducible vertex contribution and 
the non-Goldstone boson exchanges. In particular,  in \cite{BPP96,BPP02}
the one-meson irreducible vertex contribution   below some scale $\Lambda$
was identified with the ENJL quark loop contribution 
while a loop of a heavy quark  with mass $\Lambda$    
was used to  mimic the contribution massless QCD quark loop  above
$\Lambda$. The results are in Table \ref{quarkL} where one 
can see a very nice stability region when $\Lambda$ is in the interval
[0.7, 4.0] GeV.
Within the ENJL model, the ENJL quark loop is related trough Ward 
identities  to the scalar exchange which we discuss below and {\it both} 
have to be included \cite{BPP96,BPP02}.
Similar results for  the quark loop below $\Lambda$ 
were obtained in \cite{HKS98,HK02} 
though these authors didn't discuss the 
short-distance long-distance  matching.
\begin{table}
\begin{center}
\caption{Sum of the short- and long-distance 
quark loop contributions  \cite{BPP96}
as a function of the matching scale $\Lambda$.
\label{quarkL}}{
\begin{tabular}{c|cccc}
$\Lambda$ [GeV] & 0.7 & 1.0 & 2.0 &4.0\\
\hline
\rule{0cm}{13pt} $10^{10} \times a_\mu$ & 2.2 &  2.0& 1.9& 2.0
\end{tabular}}
\end{center}
\vspace*{-0.5cm}
\end{table}

  The exchange of axial-vectors and scalars in nonet symmetry
--this symmetry is exact in the large $N_c$ limit, was also included in 
\cite{BPP96,BPP02} while only the axial-vector exchange was
included in \cite{HKS98,HK02}.  The result  of the scalar
exchange obtained in \cite{BPP96} is
\be
\label{scalar}
a_\mu(\rm Scalar) = -(0.7 \pm 0.2 ) \times 10^{-10} \, .
\ee
The result of the axial-vector exchanges in \cite{HKS98,HK02}
and \cite{BPP96,BPP02} can be found in Table \ref{tab3}.
\begin{table}
\begin{center}
\caption{Results for the axial-vector exchange contributions from 
\cite{HKS98,HK02} and \cite{BPP96,BPP02}.\label{tab3}}{
\begin{tabular}{c|c}
 References
 & $10^{10} \times a_\mu$\\
\hline
 \cite{HKS98,HK02}  & 0.17 $\pm$ 0.10  \\
\cite{BPP96,BPP02}  & 0.25 $\pm$ 0.10 
\end{tabular}}
\end{center}
\vspace*{-0.5cm}
\end{table}

 Melnikov and Vainshtein used a model that saturates
the hadronic four-point function in (\ref{four}) at
leading order  (LO) in the $1/N_c$ expansion 
with $\pi^0$ and axial-vector exchanges.
In that model, the new OPE constraint of the reduced four-point
function found in \cite{MV04} forces the $\pi^0 \gamma^*(q) \gamma(p_3=0)$ 
vertex to be point-like
rather than including a ${\cal F}(q^2,0)$ 
form factor. There are also OPE constraints for
other momenta regions which are not satisfied by the model
in \cite{MV04} though they argued that this mades only a small numerical
difference of the order of $0.05 \times 10^{-10}$.  In fact, 
within the large $N_c$ framework, it has been shown \cite{BGL03}
that in general for other than two-point functions, to satisfy fully
the QCD short-distance properties requires
the inclusion of an infinite number of narrow states.

 The results in \cite{MV04}
for  the Goldstone boson exchanges and for the 
axial-vector exchanges can be found
in Table \ref{tab1} and \ref{tab3}, respectively.
\begin{table}
\begin{center}
\caption{Results quoted in Ref. \protect\cite{MV04} for
the axial-vector exchange depending of the $f_1(1285)$ 
and $f_1(1420)$ resonances mass mixing.
\label{massmixing}}{
\begin{tabular}{c|c}
Mass Mixing  & $10^{10} \times a_\mu$\\
\hline
No New OPE and Nonet Symmetry &   \\
 with M=1.3 GeV & 0.3 \\
New OPE  and Nonet Symmetry &     \\
 with M= 1.3 GeV & 0.7 \\
New OPE  and Nonet Symmetry &     \\
with M= M$_\rho$  & 2.8 \\
New OPE  and Ideal Mixing &  \\
with Experimental Masses & 2.2 $\pm$ 0.5\\ 
\end{tabular}}
\end{center}
\vspace*{-0.5cm}
\end{table}

\begin{table}
\begin{center}
\caption{Full hadronic light-by-light contribution
to $a_\mu$ at ${\cal O}(N_c)$. The difference between the
two results of Refs. \protect\cite{BPP96} and \protect\cite{BPP02}
 is the contribution of the
scalar exchange $-(0.7\pm0.1) \times 10^{-10}$.
This contribution is not included in Refs.
\protect\cite{HKS98,HK02} and  \protect\cite{MV04}.
\label{comparisontab}
\label{largeN}}{
\begin{tabular}{c|c}
Hadronic light-by-light 
at ${\cal O} (N_c)$  & $10^{10} \times a_\mu$\\
\hline
Nonet Symmetry \protect\cite{HKS98,HK02} &  9.4 $\pm$ 1.6 \\ 
Nonet Symmetry + Scalar \protect\cite{BPP96,BPP02} 
& 10.2 $\pm$ 1.9\\ 
Nonet Symmetry \protect\cite{BPP96,BPP02}&  10.9 $\pm$ 1.9  \\
 New OPE and 
Nonet Symmetry \protect\cite{MV04} &  12.1 $\pm$ 1.0 \\ 
 New OPE and 
Ideal Mixing  \protect\cite{MV04} &  13.6 $\pm$ 1.5  
\end{tabular}}
\end{center}
\vspace*{-0.5cm}
\end{table}

\section{Next-to-Leading in the $1/N_c$ Expansion Contributions} 

 At next-to-leading (NLO) in the $1/N_c$ expansion, the pion loop
is the dominant one and because the pion mass is not much larger
than the muon mass $m$, one expects a contribution 
of the order of $ 10^{-10}$.  To dress the photon 
interacting  with pions, a particular Hidden Gauge Symmetry (HGS) 
model was used in  \cite{HKS98,HK02} while  a full VMD was used 
in \cite{BPP96,BPP02}.
The results obtained are $-(0.45\pm 0.85) \times 10^{-10}$ in \cite{HKS98}
and $-(1.9\pm0.5) \times 10^{-10}$ in \cite{BPP96}.
Both models satisfy the known constraints  
though start differing at  ${\cal O}(p^6)$ in CHPT.
It is also known that the full VMD   does rather well
reproducing higher order terms of CHPT while the special
version of the HGS used in \cite{HKS98}  does not give the
correct QCD high energy behavior in some two-point functions, in particular
it does not fulfill the  Weinberg Sum Rules, see \cite{BPP96} 
for more comments.  Some studies of the cut-off dependence 
of the pion loop using the full VMD model 
was done  in \cite{BPP96}  and showed  that  their  final number comes 
from fairly low energies where the model dependence should be smaller.
The authors of \cite{MV04} analyzed the model
used in \cite{HKS98,HK02} and  showed that there is a large 
cancellation between the first three terms of an expansion in powers of 
$(m_\pi/M_\rho)^2$ and  with large higher order corrections
when expanded in CHPT orders but the same applies to the $\pi^0$
exchange as can be seen from Table 6 in the first reference in \cite{BP07} 
by comparing the WZW  column with the others.
 The authors of \cite{MV04} took 
$(0\pm 1) \times 10^{-10}$ as  a guess estimate of the total NLO in $1/N_c$
contribution. This seems too simply and  certainly with 
underestimated uncertainty.

\section{Comparison  Between Different Calculations}
 
The comparison of individual contributions 
in \cite{HKS98,HK02,BPP96,BPP02,KN02,DB08} and in \cite{MV04}   
has to be done with care because they come from different model assumptions
to construct the full relevant four-point function. 
In fact,  the authors of \cite{DB08} have shown that their
constituent quark loop  provides the correct asymptotics
and in particular the new OPE found in \cite{MV04}. 
It has more sense to compare  results for $a_\mu^{\rm lbl}$ either  at 
leading order
  or at next-to-leading order in the $1/N_c$ expansion. The recent
results for $a_\mu^{\rm lbl}$ at  LO  in the  $1/N_c$ expansion 
is what is shown in Table \ref{largeN}. 
The  nice agreement between them within the quoted uncertainty 
 leads us \cite{BP07} to take 
\be
a_\mu^{{\rm lbl}, N_c} = (11 \pm 4) \times 10^{-10}
\ee
as a robust result for the  hadronic light-by-light contribution
to muon anomaly $a_\mu$ at  LO in the $1/N_c$ expansion. 

The results for the final hadronic light-by-light 
contribution to $a_\mu$ quoted in \cite{HKS98,HK02}, \cite{BPP96,BPP02}
and \cite{MV04} are in Table \ref{tab4}.
The apparent agreement between \cite{HKS98,HK02} and \cite{BPP96,BPP02}
hides non-negligible differences which numerically almost compensate
between the quark-loop and charged pion and kaon loops. Notice also
that \cite{HKS98,HK02} didn't include the scalar exchange. 
\begin{table}
\begin{center}
\caption{Results for the full hadronic light-by-light
contribution to $a_\mu$.\label{tab4}}{
\begin{tabular}{c|c}
 Full Hadronic Light-by-Light
 & $10^{10} \times a_\mu$\\
\hline
 \cite{HKS98,HK02}  & 8.9$\pm$ 1.7  \\
\cite{BPP96,BPP02}  & 8.9 $\pm$ 3.2  \\
\cite{MV04}         & 13.6 $\pm$ 2.5 
\end{tabular}}
\end{center}
\vspace*{-0.5cm}
\end{table}
 Comparing the results of \cite{BPP96,BPP02} and \cite{MV04},
as discussed above, we have found several differences of order
$1.5 \times 10^{-10}$  which are not related to the new short-distance
constraint used in \cite{MV04}.  The different  axial-vector mass
mixing accounts for $-1.5 \times 10^{-10}$, the absence of the
scalar exchange in \cite{MV04} accounts for $-0.7 \times 10^{-10}$
and the use of a vanishing  NLO in $1/N_c$ contribution
in \cite{MV04} accounts for $-1.9 \times 10^{-10}$. These model
dependent differences add up to $-4.1 \times 10^{-10}$ out of the
final $-5.3 \times 10^{-10}$ difference between \cite{BPP96,BPP02}
and \cite{MV04}. Clearly, the new OPE constraint  used
in \cite{MV04} accounts only for a small part 
of the large numerical final difference.

\section{Conclusions}

We observe  a nice agreement, see Table \ref{largeN}, 
 between the recent model calculations of  the hadronic 
light-by-light  contribution to $a_\mu$ at LO in the $1/N_c$ expansion, 
hence  concluding  that 
\be
\label{LO}
a_\mu^{{\rm lbl}, N_c} = (11 \pm 4) \times 10^{-10}
\ee
is a very solid result.
We also understand the origin of the final numerical difference
between the results quoted in \cite{MV04} and \cite{BPP96,BPP02}.
 Its origin is not dominated by the new OPE constraint found
in \cite{MV04} and it rather comes from the addition of several 
model dependent  differences of order $1.5 \times 10^{-10}$ 
as discussed above.

 It is possible and desirable to make a new calculation 
of $a_\mu^{\rm lbl}$ using the techniques developed 
in \cite{KNP02,BGL03,CEM06} and the new OPE results \cite{MV04}.
 
The authors of \cite{PRV08} 
have  done a  conservative analysis of the
present situation of the  hadronic light-by-light contribution
to $a_\mu$ including the NLO in the $1/N_c$ expansion contribution. 

Very valuable information about various pieces of the theoretical 
models used to calculate the hadronic light-by-light contribution 
to $a_\mu$  can be obtained  by  measuring the 
$\pi^0 \to \gamma \gamma^*$, 
$\pi^0 \to \gamma^* \gamma^*$ and  $\pi^0 \to e^+ e^-$ decays which 
constrain the off-shell $\pi^0 \gamma^* \gamma^*$
and $\pi^0 \gamma^* \gamma$  form factors and the subleading 
$\mu$-dependent terms discussed in Section \ref{LOC} and by measuring 
the  $\gamma^* \gamma^* \to \pi^+ \pi^-$, $e^+ e^- \to \pi^+ \pi^-$
processes which constrain the $\pi^+ \pi^- \gamma^* \gamma^*$
vertex which dominates the uncertainty of the pion loop contribution.
 The $\gamma\gamma$ programme at the upgraded  DA$\Phi$NE-2 facility 
at Frascati is very well suited  for these measurements.

\section*{Acknowledgments}
It is a pleasure to thank Hans Bijnens,
Elisabetta Pallante, Eduardo de Rafael and Arkady Vainshtein
for enjoyable collaborations and discussions on  the different
topics    discussed here.
This work has been supported in part  by MICINN,
 Spain and FEDER, European Commission (EC) Grant No. FPA2006-05294, 
by the Spanish Consolider-Ingenio 2010 Programme CPAN 
Grant No. CSD2007-00042,  by Junta de Andaluc\'{\i}a Grants No. 
P05-FQM 101, P05-FQM 347 and P07-FQM 03048 and  by the EC RTN FLAVIAnet
 Contract No. MRTN-CT-2006-035482.

\end{document}